# Efficient Resources Distribution for an Ephemeral Cloud/Edge continuum

Emanuele Carlini
CNR-ISTI Pisa, Italy
Email: emanuele.carlini@isti.cnr.it

Patrizio Dazzi
CNR-ISTI Pisa, Italy
Email: patrizio.dazzi@isti.cnr.it

Luca Ferrucci
CNR-ISTI Pisa, Italy
Email: luca.ferrucci@isti.cnr.it

Matteo Mordacchini
CNR-IIT Pisa, Italy
Email: matteo.mordacchini@iit.cnr.it

*Abstract*—This paper presents the idea and the concepts behind the vision of an Ephemeral Cloud/Edge Continuum, a cloud/edge computing landscape that enables the exploitation of a widely distributed, dynamic, and context-aware set of resources. The Ephemeral Continuum answer to the need of combining a plethora of heterogeneous devices, which nowadays are pervasively embedding anthropic environments, with both federations of cloud providers and the resources located at the Edge. The aim of the Ephemeral Continuum is to realise a context-aware and personalised federation of computational, data and network resources, able to manage their heterogeneity in a highly distributed deployment.

## I. Introduction

Cloud Computing transformed IT by providing computing and storage as on-demand utility services according to the pay per use model. The initial concept of cloud computing has rapidly evolved into a multi-cloud environment, which gathers together multiple and heterogeneous cloud data centres and service providers. The earlier protagonist of today's multi-cloud era is the so called *Cloud Federations* [1]. In a Cloud federation, applications are submitted to a specific cloud provider. However, the execution can in principle involve any combination of providers within the federation. To achieve this result, various solutions have been advanced, ranging from centralised controllers of the federation [2] to more decentralised [3] or self-management based approaches [4].

On the other side, the devices at the edge of the network participate to a collective realisation of services, and they offload computation and data in the cloud to support these demanding applications that would have been outside their computational capacities otherwise. Nowadays, the increasing pervasiveness of Cloud-based solutions and the fast expansion of the Internet at the Edge are creating new scenarios for Cloud-based applications. In this paper we conceptualise the *Ephemeral Cloud/Edge Continuum*, a utility computing environment aimed at realising the convergence of hybrid and mobile cloud paradigms that also encompass Edge resources. In fact, the requests of a user trigger a dynamic and temporary (i.e., *ephemeral*) federation that is highly tailored to a user context and needs, depending on the user's hybrid Cloud resources and Edge resources in the surrounding environment.

## II. Challenges

Users are increasingly accessing cloud services from mobile devices in scenarios where other devices – static and mobile - available in specific physical areas, generate raw data [5], [6], [7] and provide distributed cloud computing functions. In order to fully exploit these resources, traditional cloud solutions based on remote public clouds may not necessarily be the best response in such a dynamic and localized context. In contrast, locating the computing functions towards the edge of the network has the advantage of make it easier to support services with low latency requirements and to facilitate real-time interactions between the nodes in the architecture. Therefore, the main challenge is to allow the system to optimally decide where to deploy the different parts of an application and its required data. Dynamically detecting the context of the user and the resources, including those located at the edge of the network require to face challenging issues. The Cloud/Edge continuum is a heterogeneous environment where devices may have extremely limited resources in terms of energy, storage and computing. In this case, the design of the system should consider efficient techniques that allow to face the challenge to perform service discovery in a distributed manner while ensuring low latency and fast response to network dynamics (link/node failures, mobility), as well as low energy consumption. Once the application and its required data are distributively deployed, the system should support support the execution of the service in a such a context. This fact requires to face the additional challenge of how to efficiently orchestrate in a distributed way the various service components in a system that federates public clouds as well as the user's personal private cloud and edge devices. Problems such as how to compose multiple resources, how to select the sequence of service components to execute, how to optimally migrate components and the data needed to execute a service inside this scenario should be considered.

## III. Ephemeral Cloud/Edge Continuum

To overcome the limitations affecting the current solutions there is a need for models able to encompass the edge and federated-cloud computing approaches. That is the necessity of creating a dynamic, context-aware and personalised federations of computational, data and network resources, able to manage their heterogeneity in a highly distributed deployment.

To address these needs we envision an innovative cloud approach that we identify with the term "Ephemeral Cloud/Edge Continuum". It consists in an approach allowing the deployment of cloud applications on a widely distributed and heterogeneous cloud federation infrastructure, involving both traditional cloud- and edge-resources located at the network. The approach is intended to provide mechanisms to dynamically adapt application deployment depending on the actual context (data, resources, location, etc.) and match the need for robustness, trustworthiness and performance. Starting from initial requirements on applications and data, and by exploiting monitoring and context information, the Ephemeral Cloud/Edge Continuum is able to define, enact and dynamically optimise application deployment and runtime management plans.

*A. Dynamic, Personalised and Context-aware resources*

The Ephemeral Cloud/Edge Continuum paradigm enables the exploitation of a dynamic, personalised and context-aware set of resources. In fact, the Ephemeral Cloud/Edge Continuum is said to be ephemeral as there is no a single, unique, consolidate view on the available resources as such view is personalised according to the user and depends on multiple factors. The Ephemeral Cloud/Edge Continuum meets the need for high dynamicity and high scalability, by means of a tight interaction between the orchestration of the application and the organisation and management of the resources executing the application. To this respect, the Ephemeral Cloud/Edge Continuum can represent a good fit as the personalisation inherently defines a strict relationship among application and federated resources. All these aspects concur to make the set of resources that an application can exploit very dynamic and flexible. This makes the Ephemeral Cloud/Edge Continuum a very powerful and flexible computation platform for a wide range of applications.

*B. Services on the Ephemeral Cloud/Edge Continuum*

Services are submitted to the Ephemeral Cloud/Edge Continuum along with a representation of their computational and security requirements. The Ephemeral Cloud/Edge Continuum provides an advanced support for application orchestration, allowing customers to easily drive the placement of applications and data according to their preferences and needs and matching the requirements of a robust, secure and performant infrastructure.

From an operative perspective, the allocation process is conducted by the Ephemeral Cloud/Edge Continuum at different stages, at different levels of abstraction, and at different places. Applications, their actual users and the associated contexts are analysed to derive an initial, high-level allocation plan. Such plan can be locally performed by any access point to the federation. It takes into consideration a coarse representation of all the resources available in the federation.

Basically, it consists in defining where the different instances of the services composing the application should be placed, in terms of data-centres, or overall set of resources (e.g., a certain public cloud, a certain set of edge resources). Then, it takes place a finer scheduling/mapping process targeting the specific resources selected for hosting the application, e.g. intra-cluster mapping. This completes the initial application allocation process. After, it is started the monitoring process. The purpose of this activity is to keep informed the application runtime management, devoted to the continuous optimisation and management of the applications. Such process takes into account the information deriving from the observations and monitoring of the applications at run-time, in order to provide the necessary management plans resulting in specific application re-configuration operations, such as migration as well as horizontal- and vertical-elasticity.

IV. KEY COMPONENTS

To be able to support the execution of application on a dynamic, personalised, heterogeneous and distributed set of resources, the enabling platform of the Ephemeral Cloud/Edge Continuum needs to be carefully designed and structured. In this section are presented the key building blocks characterising the platform realising the Ephemeral Cloud/Edge Continuum.

*A. Brokering and Discovery*

In the research community, there is a wide consensus on the importance that brokers can have on multi-cloud/edge environments for helping consumers in discovering, considering and comparing services with different capabilities as offered by different providers [8]. The core duties of a cloud broker [9] are to select the appropriate resource for the execution of an application. These resources satisfy the functional and non-functional requirements (including cost) expressed by the user. Due to the importance of the problem and the diffusion among industry and research of multi-cloud environments, many brokering solutions have been developed, for example exploiting rule-based algorithms [10], genetic algorithms [11] or even machine learning [12].

*B. Runtime*

Once the scheduling activity has been conducted, the service instances composing an application are properly deployed on the target resources. There exist several solutions to determine the amount of replica to create for each component of an application, both using solutions working at deployment time [13] as well at runtime [14]. It consists in an active entity that is able to ensure that the application requirements will be properly matched by exploit the information about applications, users, resources and the resulting interactions. In the context of the Ephemeral Cloud/Edge Continuum we envision a runtime support organised according to the Model-Analyse-Plan-Execute (MAPE) approach of autonomic control, relying on performance models associated to the resources and the applications that are dynamically defined and updated by leveraging the information derived from previous execution of the applications.

## C. Monitoring

An efficient monitoring subsystem is a fundamental building block for the achievement of the Ephemeral Cloud/Edge Continuum. It is involved both in supporting the resource brokering as well as in the applications scheduling and their runtime management. Along the years have been proposed many solutions for facing this aspect in the context of wide cyber-infrastructures. Both in the area of Computational Grids [15] and Clouds [16]. In fact, a proper defined monitoring support is key to define efficient and effective deployment plans. In fact, to be effective, all these activities need to be performed taking into consideration "fresh", up-to-date information. Following the best practices of large computing systems, in our envisioned scenario, the monitoring activity is conducted by two different subsystems: application monitoring and resource monitoring.

*a) Application Monitoring:* The application monitoring is conducted to observe the deployed applications. The ultimate aim is twofold; on the one side it is performed to ensure that the functional and non-functional application requirements are satisfied, and to properly react in case they would not. On the other side, it allows to monitor the application and get a feedback on its behavior, to tailor the resource previsional models and to enhance the effectiveness of the scheduling support.

*b) Resource Monitoring:* Due to the inborn nature of the Ephemeral Cloud/Edge Continuum, encompassing the hybrid-cloud model and the edge computing paradigm, requires a resource monitoring model able to discover, recruit, exploit and release different kinds of resources depending on users, applications and the actual context of usage. As matter of facts, this implies a very complex interaction among resources and applications, which to be effective needs to be driven by up-to-date and properly defined information on the current status of resources.

## D. Security and Privacy

From a security viewpoint the exploitation of cloud resources within a private cloud infrastructure is quite straightforward. However, allowing secure application orchestration within multiple or federated cloud infrastructures or in mixed infrastructures consisting of a combination of private and public resources, is far more complex. In fact, the peculiarities of the Ephemeral Cloud/Edge Continuum call for the creation of a proper security architecture. It requires advanced models for security classification, involving both cloud applications and the associated data, including definition of application security requirements through user provided security policies. This, in combination with proper enactors of security policies, at deployment time as well as at runtime will provide very effective tools for the confidentiality and integrity protection of application data, its localisation control, as well as support for encryption on edge resources.

## V. CONCLUSION

This position paper details the idea and the concepts behind the Ephemeral Cloud/Edge Continuum, an innovative multi-Cloud and Edge paradigm that enables the exploitation of a dynamic, personalised and context-aware set of computing resources. In this context, we have outlined the main challenges that arise to execute applications on a platform made of the combination of cloud datacenters and the devices at the Edge. We discussed the key enablers needed to overcome these challenges, placing such components in the overall context of the Ephemeral Cloud/Edge Continuum.